\documentstyle[amsfonts,epsfig,12pt]{article}

\input{tcilatex}

\begin{document}

\title{Mass, Action and Entropy of \ Taub-Bolt-dS Spacetimes}
\author{R. Clarkson\thanks{%
Email: rick@avatar.uwaterloo.ca}, A. M. Ghezelbash\thanks{%
Email: amasoud@sciborg.uwaterloo.ca}, R. B. Mann\thanks{%
Email: mann@avatar.uwaterloo.ca} \\
Department of Physics, University of Waterloo, \\
Waterloo, Ontario N2L 3G1, CANADA \\
Perimeter Institute for Theoretical Physics, 35 King Street North, \\
Waterloo, Ontario N2J 2W9, CANADA}
\maketitle

\begin{abstract}
We apply a recent proposal for defining conserved mass in asymptotically de
Sitter spacetimes to the class of Taub-Bolt-dS spacetimes. We compute the
action, entropy and conserved mass of these spacetimes, and find that in
certain instances the mass and entropy can exceed that of pure de Sitter
spacetime, in violation of recent suggestive conjectures to the contrary.
\end{abstract}

The search for a holographic dual to de Sitter spacetime has led to a number
of suggestive results and conjectures concerning gravitational
thermodynamics and its underlying quantum-gravitational description.
Intriguing evidence that de Sitter (dS) spacetime has maximal entropy first
came from calculations of cosmological black hole pair production \cite%
{simonivan} and was recently formulated in terms of the N-bound \cite{bousso}%
, which states that any spacetime with positive cosmological constant $%
\Lambda =3\ell ^{-2}$ has observable entropy $S\leq S_{\text{dS}}=\pi \ell
^{2}$. From this followed the notion that dS spacetime also had maximal
mass/energy, expressed in the form of a conjecture \cite{bala} that {\it any
asymptotically dS spacetime with mass greater than pure dS has a
cosmological singularity. }

The physical meaning of a conserved mass/energy outside the cosmological
horizon of an asymptotically dS spacetime requires clarification \cite{Shiro}%
, since asymptotically dS spacetimes do not have spatial infinity the way
that their flat or anti de Sitter (AdS) counterparts do. Moreover one cannot
define a global timelike Killing vector; rather there is a timelike Killing
vector field $\partial /\partial t$ inside the dS cosmological horizon that
becomes spacelike outside of it. Recently, however, a proposal to extend the
Brown-York prescription to asymptotically dS spacetime \cite%
{brown,BCM,ivan,balakraus} yielded suggestive information about the stress
tensor and conserved charges of the hypothetical dual Euclidean conformal
field theory (CFT) on their spacelike boundaries, providing intriguing
evidence for a holographic duality similar to the AdS/CFT correspondence.
The specific prescription in ref. \cite{bala} (see also \cite{Klem, Myung})
introduced counterterms on spatial boundaries at past and future infinity
(algorithmically derivable from the Gauss-Codacci equation \cite{GM}) that
yielded a finite action for asymptotically dS spacetimes. Carrying out a
procedure analogous to that in the AdS case \cite{BCM,balakraus}, one
obtains a stress tensor on the past/future boundary, and consequently a
conserved quantity associated with each Killing vector there.

Interpreting the mass/energy as the charge associated with the extension of $%
\partial /\partial t$ outside the cosmological horizon leads to the above
conjecture, where the cosmological singularity is a timelike region of
geodesic incompleteness at which scalar Riemann curvature invariants
diverge; the negative-mass Schwarzschild de Sitter solution provides a
paradigmatic example. This conjecture has been further verified for
topological dS solutions and its dilatonic variants \cite{cai} and for
Schwarzschild-de Sitter (SdS) black holes up to nine dimensions \cite{GM}.

Here we demonstrate that (locally) asymptotically dS spacetimes with NUT
charge provide counterexamples to this conjecture under certain
circumstances. Such spacetimes do not admit spin structures and so can
provide useful information as to the limitations of holographic duality. We
find that these spacetimes can have a positive mass whilst respecting the
N-bound on entropy. While we treat only the four-dimensional case, our
results can be straightforwardly extended to arbitrary (even) dimensions.
Extending the notion of entropy to past/future infinity \cite{GM} we also
obtain violations of the N-bound in certain instances, though its
application to this situation is in need of further clarification.

NUT-charged dS spacetimes are exact solutions to the Einstein equations and
have metrics of the form 
\begin{equation}
ds^{2}=V(\tau )(dt+2n\cos \chi d\phi )^{2}-\frac{d\tau ^{2}}{V(\tau )}+(\tau
^{2}+n^{2})(d\chi ^{2}+\sin ^{2}\chi d\phi ^{2})  \label{TBDS}
\end{equation}%
where 
\begin{equation}
V(\tau )=\frac{\tau ^{4}+\left( 6n^{2}-\ell ^{2}\right) \tau ^{2}+2m\ell
^{2}\tau -n^{2}\left( 3n^{2}-\ell ^{2}\right) }{\left( \tau
^{2}+n^{2}\right) \ell ^{2}}  \label{VV}
\end{equation}%
with $n$ \ the nonvanishing NUT charge, which can be regarded as a magnetic
type of mass. The coordinate $t$ parametrizes a circle fibered over the
non-vanishing sphere parametrized by $(\chi ,\phi )$ and must have
periodicity$\left| \frac{8\pi n}{k}\right| $ to avoid conical singularities,
where $k$ is a positive integer. The geometry of a constant-$\tau $ surface
is that of a Hopf fibration of $S^{1}$ over $S^{2}$, and the metric (\ref%
{TBDS}) describes the contraction/expansion of (for $k=1$) this 3-sphere in
spacetime regions where $V(\tau )>0$ outside of the past/future cosmological
horizons. No closed timelike curves exist in these regions, and the
spacetime is asymptotically locally de Sitter, with $\left| k\right| $
points identified around the $S^{1}$ fibres.

The spacelike Killing vector $\partial /\partial t$\ has a fixed point set
where $V(\tau _{0})=0$\ whose topology is that of a 2-sphere. Since $\frac{%
\partial }{\partial \phi }$\ is a Killing vector, for any constant $\phi $
-slice near the horizon $\tau =\tau _{0}$\ additional conical singularities
will be introduced in the $(t,\tau )$\ Euclidean section unless $t$\ has
period $4\pi /\left| V^{\prime }\left( \tau _{0}\right) \right| $. This
period must equal $\left| \frac{8\pi n}{k}\right| $, which forces $\tau
_{0}=\tau ^{\pm }$\ where 
\begin{equation}
\tau ^{\pm }=\frac{k\ell ^{2}\pm \sqrt{k^{2}\ell ^{4}-144n^{4}+48n^{2}\ell
^{2}}}{12n}  \label{taus}
\end{equation}%
yielding two distinct extensions\ (TB$^{\pm }$) to dS spacetime called
Taub-bolt-de Sitter space because of the dimensionality of the fixed-point
set of $\partial /\partial t$. The respective mass parameters are%
\begin{equation}
m^{\pm }=-\frac{k^{3}\ell ^{6}\pm (288n^{4}-24\ell ^{2}n^{2}+k^{2}\ell ^{4})%
\sqrt{k^{2}\ell ^{4}-144n^{4}+48n^{2}\ell ^{2}}}{864n^{3}\ell ^{2}}
\label{massesonn}
\end{equation}%
where 
\begin{equation}
\left| n\right| \leq \frac{1}{6}\ell \sqrt{6+3\sqrt{4+k^{2}}}  \label{nmax}
\end{equation}%
so that $\tau ^{\pm }$ are both real. Without loss of generality we can take 
$n>0$; results for $n<0$ can be obtained by reversing the signs of $t$ and $%
\phi $. Note that TB$^{-}$\ does not exist for $\left| n\right|
<n_{c}=.2658\,\ell $\ since $V^{-}(\tau )$ then develops two additional
larger real roots, and the periodicity condition cannot be satisfied.

The spacetime (\ref{TBDS}) is free of scalar curvature singularities -- the
invariants $R_{\mu \nu \rho \lambda }R^{\mu \nu \rho \lambda }$\ and $\sqrt{%
-g}R_{\mu \nu }^{\quad \,\alpha \beta }\epsilon _{\alpha \beta \rho \lambda
}R^{\mu \nu \rho \lambda }$ are finite everywhere. However NUT-charged
spacetimes in general have quasiregular singularities (end points of
incomplete and inextendable geodesics at which the Riemann tensor and its
derivatives remain finite in all parallelly propagated orthonormal frames) %
\cite{Konko}. We interpret the cosmological singularities in the conjecture
of ref. \cite{bala} to imply that scalar Riemann curvature invariants will
diverge to form timelike regions of geodesic incompleteness whenever the
conserved mass of a spacetime becomes positive (i.e. larger than the zero
value of pure dS).

To compute the conserved mass, we consider the $\left( d+1\right) $
dimensional action that yields the Einstein equations with a positive
cosmological constant 
\begin{equation}
I=\frac{1}{16\pi }\int_{{\cal M}}d^{d+1}x\sqrt{-g}\left( R-2\Lambda +{\cal L}%
_{M}\right) -\frac{1}{8\pi }\int_{{\cal \partial M}^{-}}^{{\cal \partial M}%
^{+}}d^{d}x\sqrt{h^{\pm }}K^{\pm }+I_{ct}  \label{totaction}
\end{equation}%
where ${\cal \partial M}^{\pm }$ are the future/past boundaries, and $\int_{%
{\cal \partial M}^{-}}^{{\cal \partial M}^{+}}d^{d}x$ indicates an integral
over a future boundary minus an integral over a past boundary, with
respective induced metrics $h_{\mu \nu }^{\pm }$ and intrinsic/extrinsic
curvatures $K_{\mu \nu }^{\pm }$ and $\hat{R}\left( h^{\pm }\right) $
induced from the bulk spacetime metric $g_{\mu \nu }$. ${\cal L}_{M}$ refers
to the matter Lagrangian, which we shall not consider here. $I_{ct}$ is the
counter-term action, calculated to cancel the divergences from the first two
terms (given in \cite{GM}) and we set $G=1$. The associated boundary
stress-energy tensor is obtained by the variation of the action with respect
to the boundary metric, the explicit form of which can be found in \cite{GM}.

If the boundary geometries have an isometry generated by a Killing vector $%
\xi ^{\pm \mu }$, then it is straightforward to show that $T_{ab}^{\pm }\xi
^{\pm b}$ is divergenceless, from which it follows that there will be a
conserved charge ${\frak Q}^{\pm }$ between surfaces of constant $t$, whose
unit normal is given by $n^{\pm a}$. Physically this means that a collection
of observers on the hypersurface all observe the same value of ${\frak Q}$
provided this surface had an isometry generated by $\xi ^{b}$ (for explicit
calculations, see{\it \ }\cite{GM}). If $\partial /\partial t$ is itself a
Killing vector, then this can be defined as ${\frak Q}^{\pm }={\frak M}^{\pm
}$, the conserved mass associated with the future/past surface $\Sigma ^{\pm
}\left( \tau \right) $ at any given point $t$ on the boundary. This quantity
changes with the cosmological time $\tau $. Since all asymptotically de
Sitter spacetimes must have an asymptotic isometry generated by $\partial
/\partial t$, there is at least the notion of a conserved total mass ${\frak %
M}^{\pm }$ for the spacetime in the limit that $\Sigma ^{\pm }$ are
future/past infinity.

\begin{figure}[tbp]
\begin{center}
\epsfig{file=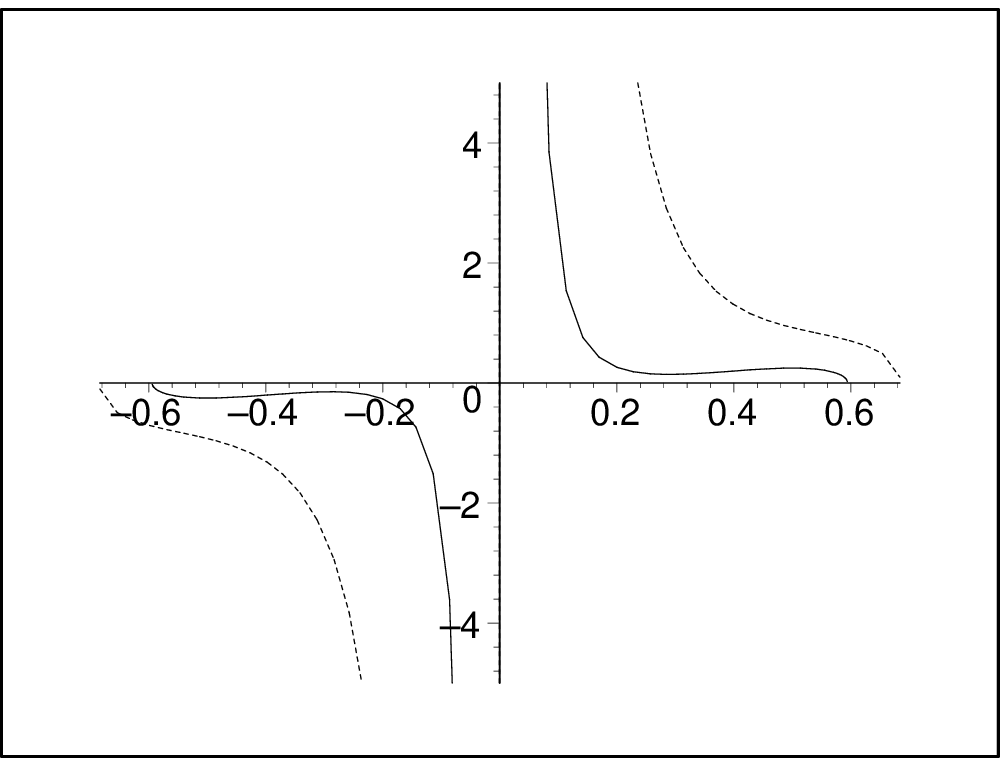,width=0.3\linewidth} %
\epsfig{file=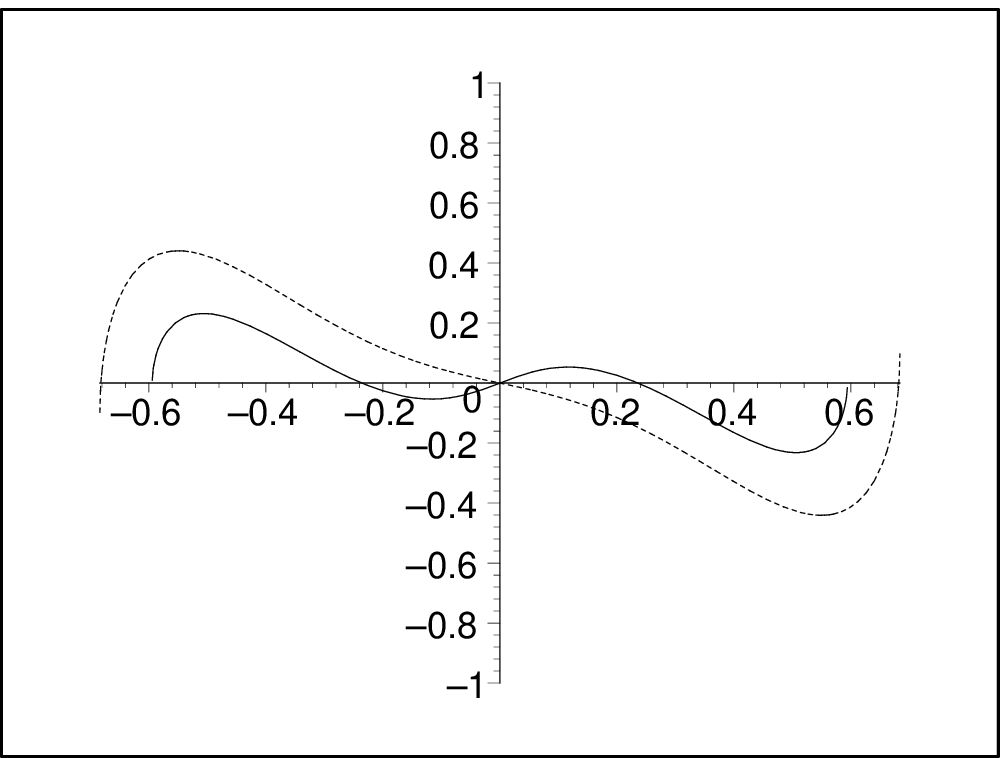,width=0.3\linewidth}
\end{center}
\caption{Left/right: Total mass ${\frak M}^{+/-}{\frak /}{\Bbb \ell }$ of
the TB$^{+/-}$ solution versus $n/\ell $ at future infinity, for $k=1$
(solid) and $k=3$ (dotted)$.$ The spacetimes TB$^{-}$ with $k=1$ and $k=3$
exist for $\left| n\right| >n_{c}\simeq .2658$ and $\left| n\right|
>n_{c}=.1879$ respectively$.$ }
\label{fig1}
\end{figure}

We obtain 
\begin{equation}
{\frak M}=-m+\frac{105n^{4}-30n^{2}\ell ^{2}+\ell ^{4}}{8\ell ^{2}\tau }+O(%
\frac{1}{\tau ^{2}})  \label{totmass}
\end{equation}%
for the metric (\ref{TBDS}) near future infinity, which for $n=0$ reduces
exactly to the total mass of the $d=4$ Schwarzschild-dS black hole \cite{GM}%
. Insertion of (\ref{massesonn}) yields this value explicitly for TB$^{\pm }$%
. Figure (\ref{fig1}) illustrates the behavior of ${\frak M}^{\pm }/\ell $
at future infinity, as a function of $n/\ell $ for $k=1$ and $3$.

While TB$^{-}$ has ${\frak M}^{-}<0$ for positive $n>n_{c}$ ($n_{c}$ depends
on $k$ and by increasing $k,$ it reduces to zero), TB$^{+}$\ has ${\frak M}%
^{+}>{\frak M}^{\text{dS}}=0$ and so forms a class of spacetimes that are
counterexamples to the conjecture proposed in \cite{bala} , since ${\frak M}%
^{+}>0$\ and there are no cosmological singularities. The signs of ${\frak M}%
^{\pm }$ are reversed for negative $n$, in which case TB$^{-}$ violates the
conjecture. 

The total action (\ref{totaction}) of the TB$^{\pm }$ spacetimes is 
\begin{equation}
I^{\pm }=-\frac{\beta _{H}}{2\ell ^{2}}(m^{\pm }\ell ^{2}+\left( \tau ^{\pm
}\right) ^{3}+3n^{2}\tau ^{\pm })  \label{acttot}
\end{equation}%
where $\beta _{H}=\left| \frac{-4\pi }{V^{\prime }(\tau )}\right| _{\tau
=\tau _{0}}=$ $\left| \frac{8\pi n}{k}\right| $ is the analogue of the
Hawking temperature outside of the cosmological horizon. The parameters $%
m^{\pm }$ and $\tau ^{\pm }$ in (\ref{acttot}) are given by (\ref{massesonn}%
) and (\ref{taus}).

We turn next to an evaluation of the entropy via the relation $S=\lim_{\tau
\rightarrow \infty }\left( \beta _{H}{\frak M}-I\ \right) $\ , extending the
usual definition to asymptotically dS cases \cite{GM}. It has been shown for 
$\left( d+1\right) $-dimensional SdS space that $S$ is a positive
monotonically increasing function of ${\frak M}$, and that in $(2+1)$
dimensions is consistent with the Cardy formula \cite{bala,GM}, {\Large \ }%
providing suggestive evidence in favour of the second law in this context.%
{\Large \ }In the present case we obtain

\begin{equation}
S^{\pm }=-\frac{\beta _{H}}{2\ell ^{2}}(m^{\pm }\ell ^{2}-\left( \tau ^{\pm
}\right) ^{3}-3n^{2}\tau ^{\pm })  \label{S}
\end{equation}%
and it is straightforward to show that the first law $dS^{\pm }=\beta _{H}d%
{\frak M}^{\pm }$ is satisfied for TB$^{\pm }$ respectively.\ Figure \ref%
{fig2} shows the behavior of ${\cal S}^{\pm }=S^{\pm }/\ell ^{2}$ as a
function of $n/\ell $ for $k=1,3$.

For $k=1$ the entropy ${\cal S}^{-}\left( n\right) =-{\cal S}^{-}\left(
-n\right) $ attains for increasing $n$ a positive maximum near $n=0.116$,
vanishes near $n=0.175$, attains a negative local minimum at $n=0.506$,
increasing again to positive values for $n>0.590$ before attaining a global
positive maximum at the critical bolt charge $n_{c}=\frac{1}{6}\sqrt{6+3%
\sqrt{5}}\approx 0.594$. For all values of bolt charge the N-bound \cite%
{bousso} on the entropy ( ${\cal S}^{-}\leq $ $\pi $ ) is satisfied. For the
other values of $k$, the N-bound on the entropy also is satisfied. So for TB$%
^{-}$ and $k=1$, the N-bound is satisfied everywhere while the mass
conjecture holds only for positive bolt charge $n>n_{c}.$

Similar considerations for TB$^{+}$ imply that for positive bolt charge $n$
and any value of $k,$ both the N-bound and the mass conjecture are violated.
\ Indeed, the quasi-regular singularity structure of the spacetime is
unaltered for any choice of these parameters. \ \bigskip 
\begin{figure}[tbp]
\begin{center}
\epsfig{file=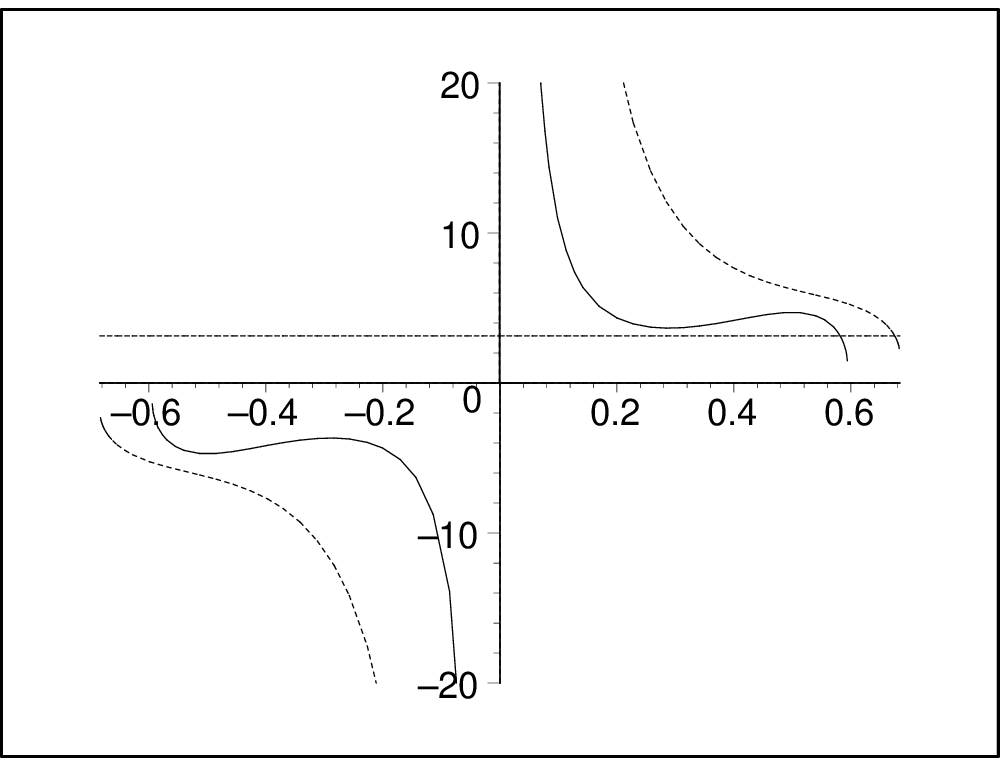,width=0.3\linewidth} %
\epsfig{file=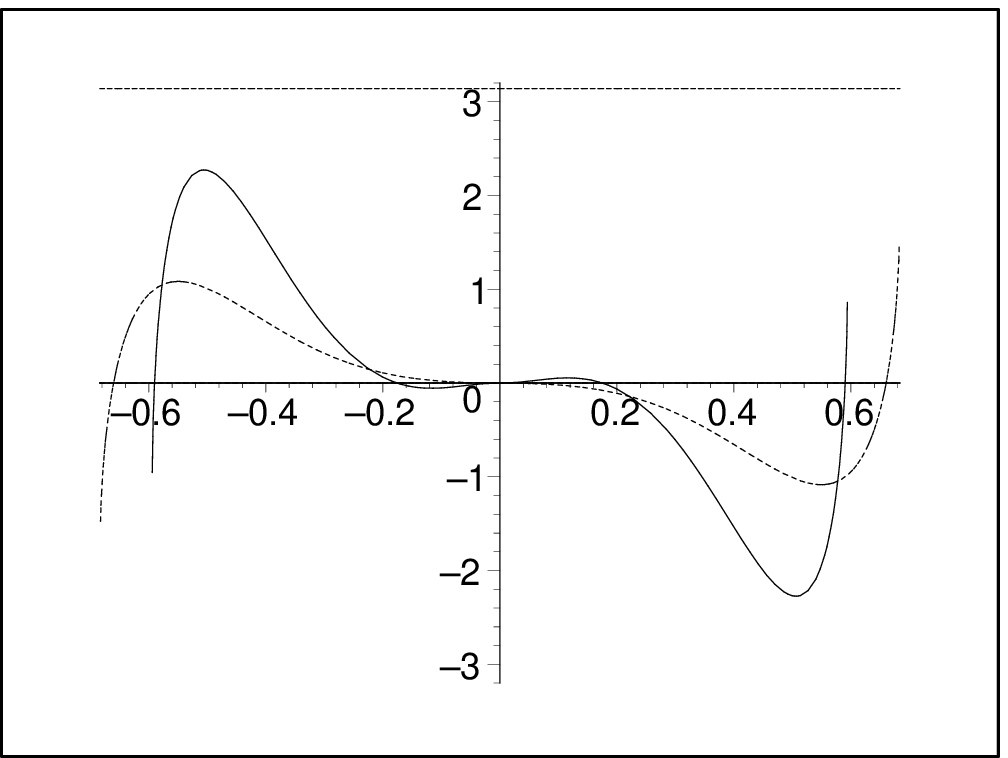,width=0.3\linewidth}
\end{center}
\caption{Left/right : Entropy ${\cal S}^{+/-}$ of TB$^{+/-}$ versus $n/\ell $
with $k=1$ (solid) and $k=3$ (dotted). The horizontal dashed lines denote
the N-bound.}
\label{fig2}
\end{figure}

Since the usual relationship between entropy and area does not hold for
NUT-charged spacetimes \cite{NUTrefs} it is natural to inquire if the
horizon area of dS spacetime is maximal. We find that while the (fixed-$t$)
area of \ the cosmological horizon of TB$^{-}$ is always less than that of
pure dS spacetime, that of TB$^{+}$ exceeds it for $n<0.2425$.

These results suggest some possible limitations to the application of the
holographic conjecture to spacetimes with $\Lambda >0$. \ For example, one
implication of the N-bound (and its associated mass conjecture) is that
spacetimes with $\Lambda >0$ cannot be described by a quantum gravity theory
with an infinite number of degrees of freedom (such as M-theory) \cite%
{bousso}. Our results suggest that this obstruction is not necessarily an
obstruction in principle, but can be overcome in spacetimes with NUT charge
that are locally asymptotically dS. \ Of course such spacetimes contain
causality-violating regions with closed timelike curves, and one might wish
to restrict their appearance in the spectrum of states of quantum gravity. \
The mechanism for so doing remains an unsolved problem.

\medskip

{\Large Acknowledgments}

This work was supported by the Natural Sciences and Engineering Research
Council of Canada. We are grateful to R. Bousso for helpful correspondence.

\end{document}